# $^{59}$Co-NMR Knight Shift of Superconducting $Na_xCoO_2 \cdot yH_2O$

Yoshiaki Kobayashi, Mai Yokoi and Masatoshi Sato

*Department of Physics, Division of Material Science, Nagoya University, Furo-cho, Chikusa-ku Nagoya, 464-8602 Japan*

Layered Co oxide $Na_xCoO_2 \cdot yH_2O$ with the superconducting transition temperature $T_c$ =4.5 K has been studied by $^{59}$Co-NMR. The Knight shift $K$ estimated from the observed spectra for powder sample exhibits almost temperature($T$)-independent behavior above $T_c$ and decreases with decreasing $T$ below $T_c$. This result and the existence of the coherence peak in the spin-lattice-relaxation-rate *versus* $T$ curve reported by the present authors indicate, naively speaking, that the singlet order parameter with $s$-wave symmetry is realized in $Na_xCoO_2 \cdot yH_2O$. Differences of the observed behaviors between the spectra of the non-aligned sample and the one aligned in epoxy adhesive by applying the external magnetic field are discussed.

corresponding author: M. Sato (e-mail: e43247@nucc.cc.nagoya-u.ac.jp)

The recent discovery of superconductivity in the layered cobalt oxide $Na_xCoO_2{\cdot}yH_2O$ [1)] has attracted much attention, because it has the triangular lattice of Co atoms. The system can be obtained fom $Na_xCoO_2$ by deintercalating $Na^+$ ions and simultaneous intercalating $H_2O$ molecules. The superconducting transition temperature $T_c$ was reported to be ~5 K. Up to now, various kinds of experimental and theoretical studies have been carried out to investigate the mechanism of the superconductivity[1-16)].

In the previous report, the present authors have shown by $^{59}$Co-NQR studies of the system, that the coherence peak of the $^{59}$Co-nuclear spin relaxation rate $1/T_1$ exists just below $T_c$. It has been also pointed out, based on detailed comparison of the relaxation rates $1/T_1$ and the magnetic susceptibilities $\chi$ between $Na_xCoO_2{\cdot}yH_2O$ and $Na_xCoO_2$ that the former system is closer to the ferromagnetic phase than the latter[7)].

In the present paper, the temperature ($T$) dependence of Knight shift $K$ studied by $^{59}$Co-NMR of $Na_xCoO_2{\cdot}yH_2O$ is reported. Then, based on the observed behaviors of $K$ and $1/T_1$, the paring symmetry is argued.

The powder sample of $Na_xCoO_2{\cdot}yH_2O$ with $T_c$ ~4.5 K was prepared from $Na_{0.75}CoO_2$ by the method described in Ref. 1. The structural and magnetic properties of the sample were reported in the previous paper[7)]. The NMR experiments were carried out using a standard phase coherent type pulse spectrometer. $^{59}$Co NMR spectra were obtained by recording the spin-echo intensity with the applied magnetic field changed step by step in a magnetic field range of 1-2 T. Measurements were carried out on two kinds of samples of $Na_xCoO_2{\cdot}yH_2O$, randomly oriented and aligned samples. The latter sample was prepared by mixing the $Na_xCoO_2{\cdot}yH_2O$ powder with epoxy adhesive (Stycast 1266) and keeping the mixture in a magnetic field of 11T at room temperature for ~12 h Its X ray diffraction pattern indicates that the *ab* plane of $Na_xCoO_2{\cdot}yH_2O$ crystallites align along the direction of the magnetic field with the spreading of about (±6°). ( The lattice parameter *c* was found not to exhibit significant change after the alignment processes, which indicates no drastic change of the structure.) The former sample does not have epoxy adhesive. It was just put into a cylindrical holder made by cellophane tapes. The obtained spectra are shown in Fig. 1(a) and 1(b) for randomly oriented powder and aligned samples, respectively.

The positions of the peaks and shoulders of the $^{59}$Co-NMR spectra observed for both kinds of samples were well explained by considering the anisotropic Knight shifts $K_x \neq K_y \neq K_z \neq K_x$ and the effects of electric quadrupole interaction up to the second order. The directions of the principal axes of Knight shift are assumed to be same as those of the electric quadrupole interaction. In the present case, the $z$-axis was defined along the direction, for which the component of the electric quadrupole tensor, $\nu_{zz}$ is larger than the other components. This $z$ direction was found to be along the $c$-axis by fitting the positions of calculated spectra to those of aligned sample. The fitted result for the data taken at 5 K in the magnetic field $\boldsymbol{H}$ within the *ab* plane of $Na_xCoO_2{\cdot}yH_2O$ is shown in Fig. 1(b) and the parameters determined by the fitting are, $K_x$= ~3.5±0.1 %, $K_y$= ~3.1±0.05 %, the electric quadrupole resonance frequency $\nu_Q$ =3.6±0.05 MHz and the asymmetric parameter of the electric field gradient at nuclei $\eta$=0.28±0.02. For the spectra of the randomly oriented sample, $K_x$= ~3.8±0.1 %, $K_y$= ~3.14±0.05 %, $K_z$= ~2.3±0.2 %, $\nu_Q$ =4.05±0.05 MHz and $\eta$=0.24±0.01 were found to reproduce them well. The comparison of the values of $\nu_Q$ and $\eta$ of the aligned sample with those of the randomly oriented one justifies the present assignment of the *z* direction along the *c* axis.

Figure 2 shows the *T* dependence of the $^{59}$Co-NMR spectra around the peak of central line for the randomly oriented powder sample. The *H* value of the peak position is nearly *T*-independent above $T_c$ (~4.5 K) and then increases with decreasing *T* below $T_c$. This change of the peak position is due to the *T* dependence of the Knight shift $K_y$ and not due to the second order effect of the electrical quadrupole interaction, because $\nu_Q$ and $\eta$ are *T*-independent, as was confirmed by analyzing the spectra in the *H* region between 0.8-2.4 T at 2.9 K, 4.7 K and 6.2 K.

It should be noted here following facts. For the aligned sample, the peak position of the central line of the $^{59}$Co-NMR spectra, which corresponds to $K_y$, is nearly *T*-independent even below $T_c$. It is slightly puzzling, because the result indicates that the *T* dependence of $K_y$ of the aligned sample is different from that of the powder sample. To explain this puzzle, we consider that the shift $K_y$ observed in ref. 8 for a sample aligned in epoxy adhesive (Stycast) also exhibits the *T*-independent behavior. We think that the specimens embedded in the Stycast have somewhat different properties from those of the (non-aligned) samples. This idea is supported by the fact that we have observed different values of $K_x$ between the two kinds of samples, which has already been stated above. Then, the data of the aligned sample may not be intrinsic.

We have estimated the $K_y$ value of the (non-aligned) powder sample. The results are shown in Fig. 3. Above $T_c$, $K_y$ is nearly constant and decreases with decreasing *T* below $T_c$. Because we have already taken into account the effect of the electrical quadrupole interaction up to the second order in the estimation of $K_y$, the decrease of $K_y$ below $T_c$ with decreasing *T* should be due to the *T* dependence of its spin component $K_{spin,y}$ and/or due to the effect of the shielding diamagnetism (The orbital component $K_{orb,y}$ may be *T*-independent.) Here, in order to estimate the change of the magnetic field at the Co nuclei by the shielding supercurrent, we consider the case where the *H* is applied along the *y* direction, i.e., within the *a*-*b* plane of the crystallites, because we are observing $K_y$. For the penetration depth $\lambda$ of 5000~8000 Å at low temperatures,[3,6)] we expect that the reduction of the field is roughly one order of magnitude smaller that the observed shift of the peak. It has been confirmed by the following experimental observation. Even when the measuring frequency is changed from 16.09 MHz to 30.044 MHz, the Knight shift does not change within the experimental error bars, indicating that the spatial variation of the field or the shielding effect is negligible in the present experimental condition. It is also confirmed by studying experimental observations for the similar two-dimensional system $YBa_2Cu_3O_7$ with $\lambda$ (~1400 Å at $T \ll T_c$) larger than that of the present system does not have a serious effect of the shielding diamagnetism in the vortex state (at $H$=7.4 T).[17)] Furthermore, it can be noted that the shape of the peak corresponding to $K_y$ does not exhibit an appreciable change expected in the case where the spatial variation of $H$ induced by the shielding diamagnetism is a main cause of the shift.

We have just tried to fit the $K_y$-$T$ curve below $T_c$ by the Yosida function[18)] (see the solid curve in Fig. 3) and obtained the superconducting energy gap $\Delta$ of 8.1 K. For the value and $T_c$ =4.5 K, $2\Delta/k_BT_c$ is estimated to be 3.6, which agrees well with the one expected from the BCS theory, though the rather good fitting by the Yosida function may be accidental. To estimate $K_{spin,y}$ precisely, we have to know the orbital component $K_{orb,y}$, which is, at this moment, not easy.

Now, we have shown that the spin susceptibility $\chi_{spin}$ of $Na_xCoO_2{\cdot}yH_2O$ studied by measuring the Knight shift $K_y$ decreases with decreasing $T$ in the superconducting state. The existence of the coherence peak of the nuclear relaxation rate $1/T_1$ just below $T_c$ has also been reported in the previous paper by the present authors.[7)] These results indicate, naively speaking, that the superconducting pairs of $Na_xCoO_2{\cdot}yH_2O$ are in the singlet spin state with $s$-wave symmetry, though other possibilities may not be completely ruled out: Strictly speaking, the ($d_1$+$id_2$) state predicted in refs. 10-13 may have a small coherence peak. Then, the precise calculation of the amplitude of the coherence peak is required to exclude such the symmetry. Even for the $p$-wave state predicted in refs. 14 and 15, the decrease of $K_{spin,y}$ can be expected, if the direction of the triplet spins is pinned in the direction perpendicular to the $y$-axis. At this moment, we do not know how strong the pinning force is. To answer this question, further studies on the anisotropy of the Knight shift has to be carried out by using single crystal specimens.

In summary, the $T$ dependence of the Knight shift $K_y$ of $Na_xCoO_2{\cdot}yH_2O$ has been reported. For the (non-aligned) powder sample, the $T$ dependence of the spin component $K_{spin,y}$ roughly follows the Yosida function in the superconducting state. This behavior and the existence of the coherence peak in the $1/T_1$-$T$ curve observed just below $T_c$ indicate that $Na_xCoO_2{\cdot}yH_2O$ has the spin singlet pairs with the $s$-wave symmetry, though the observed results do not completely exclude the possibility of other kinds of electron pair state.

Acknowledgments-The authors thank Prof. Y. Ono of Nagoya University for fruitful discussion. The work is supported by Grants-in-Aid for Scientific Research from the Japan Society for the Promotion of Science (JSPS) and by Grants-in-Aid on priority area from the Ministry of Education, Culture, Sports, Science and Technology.

Figure captions

Fig. 1(a) $^{59}$Co-NMR spectra of $Na_xCoO_2 \cdot yH_2O$ taken at $T$ =4.72 K and with the resonance frequency $f$ =16.09 MHz for the randomly oriented sample. Solid line is just for the guide to the eye.

Fig. 1 (b) $^{59}$Co-NMR spectra of $Na_xCoO_2 \cdot yH_2O$ taken at $T$ =4.72 K and with the resonance frequency $f$ =16.09 MHz for the aligned sample, where the powder of $Na_xCoO_2 \cdot yH_2O$ is embedded in the epoxy adhesive (Stycast 1266). Solid line is just for the guide to the eye. Broken line shows the spectra calculated with the parameters $K_x$=~3.5 %, $K_y$= ~3.1 %, $\nu_Q$ =3.6 MHz and $\eta$=0.28.

Fig. 2 Profiles of the sharp peak of the central line taken for the randomly oriented powder sample at several fixed temperatures are shown. The peak positions correspond to the Knight shifts $K_y$. The vertical broken line shows the peak position above $T_c$.

Fig. 3 Knight shift $K_y$ is plotted against $T$. The solid line shows the Yosida function fitted to the data.

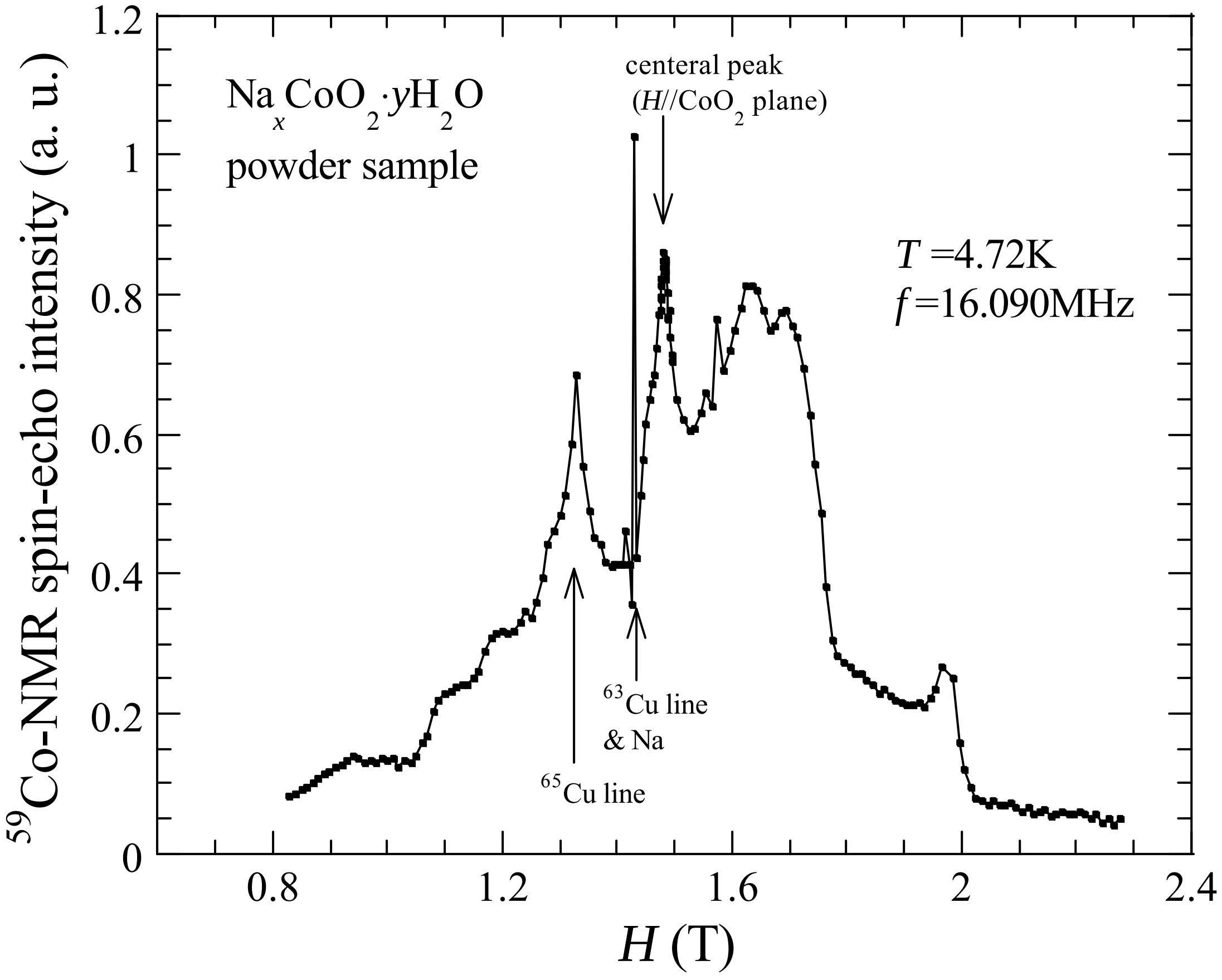

Fig. 1(a)

Y. Kobayashi *et al*.

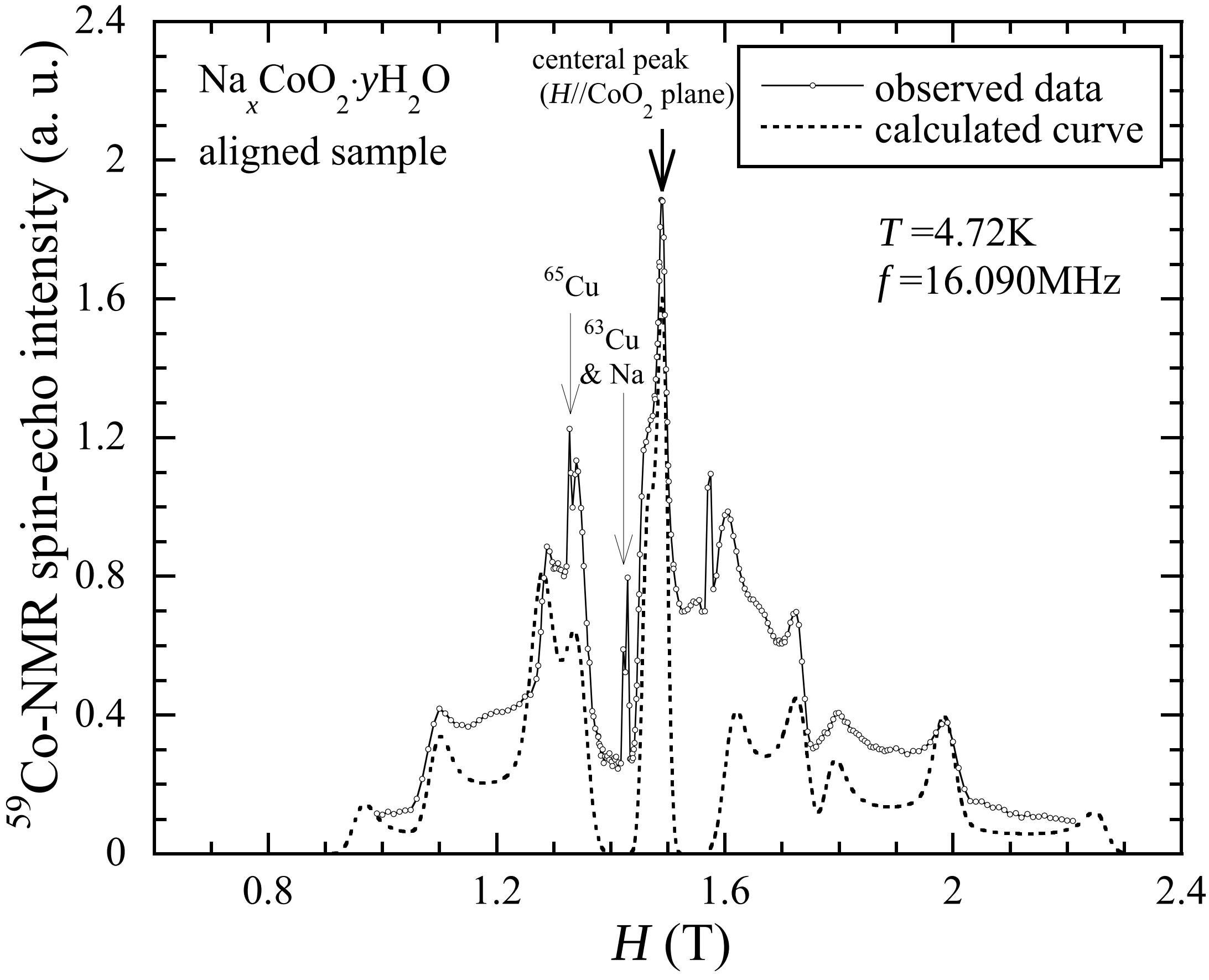

Fig. 1(b)

Y. Kobayashi *et al.*

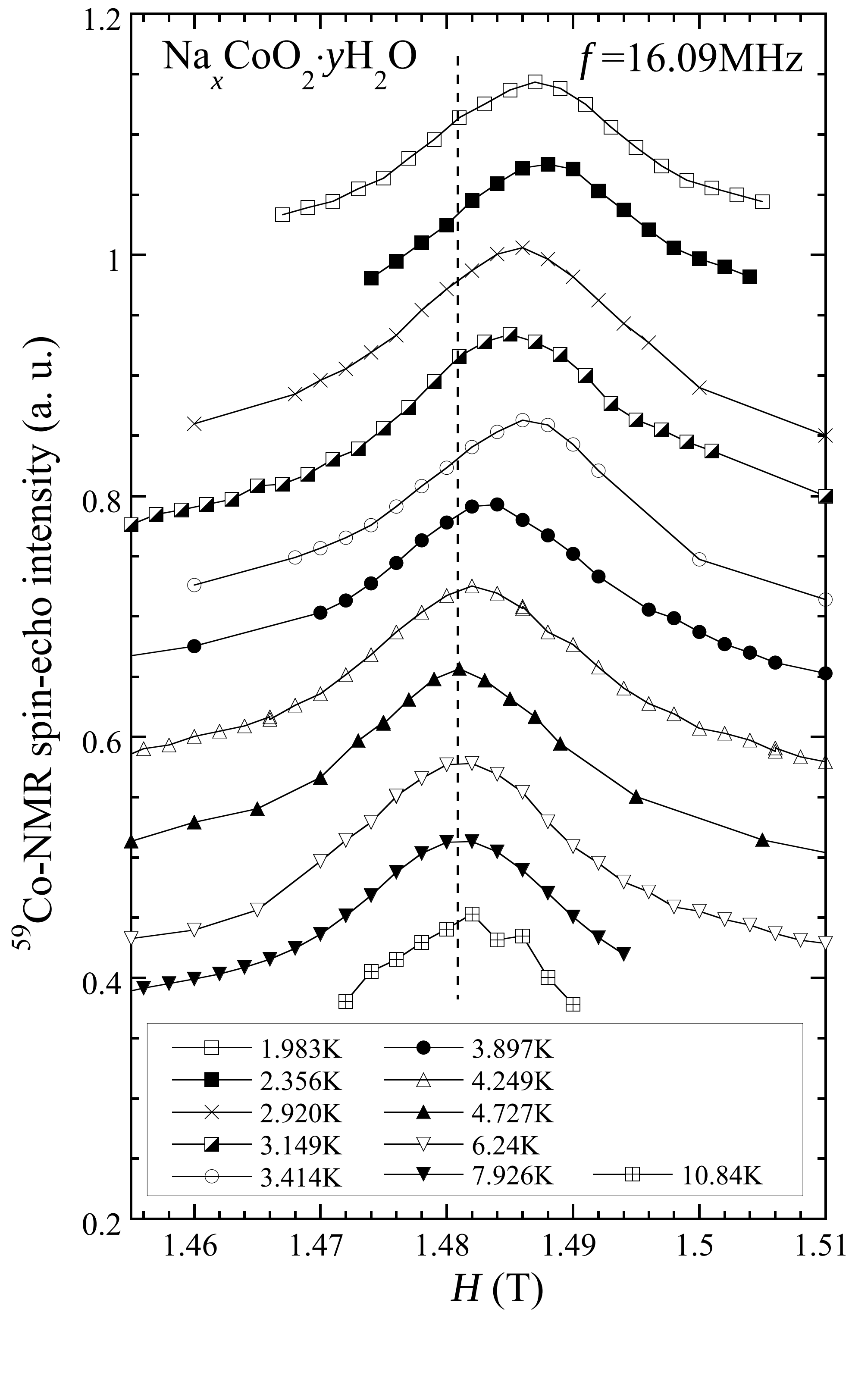

Fig. 2

Y.Kobayashi *et al.*

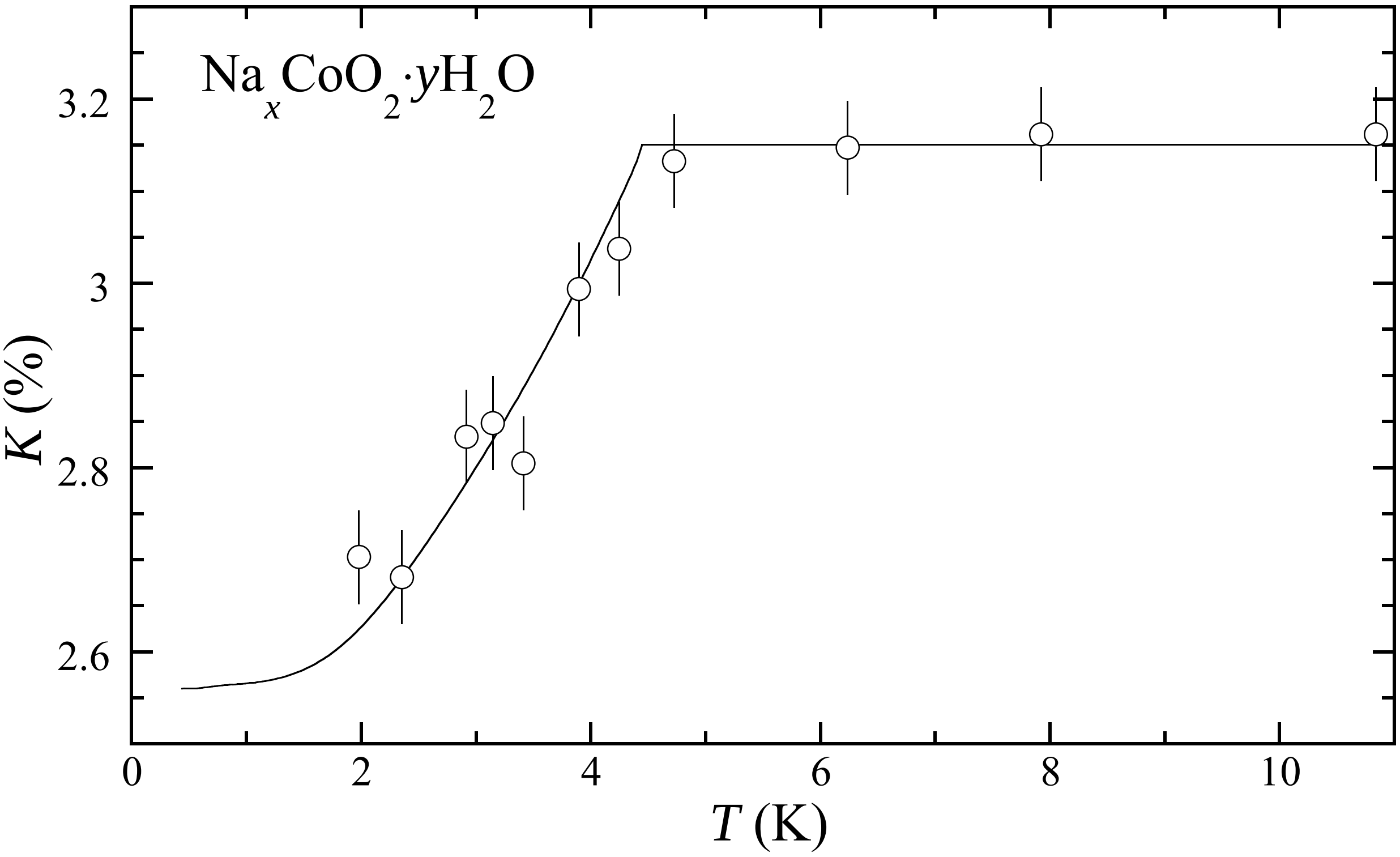

Fig. 3

Y. Kobayashi *et al.*